\documentclass[reprint,aps,prapplied,longbibliography,twocolumn,10pt,superscriptaddress,nobibnotes]{revtex4-2}

\usepackage{graphicx}
\usepackage{dcolumn}
\usepackage{bm}
\usepackage{hyperref}
\usepackage{soul}
\hypersetup{%
	colorlinks = true,
	citecolor = blue,
	linkcolor = blue,
	urlcolor = blue,
	linkbordercolor = 0 0 0,
	pdfborder= 0 0 0,
}

\usepackage[pagewise, mathlines]{lineno}
\usepackage{physics}
\usepackage{xcolor}

\usepackage{multirow}

\definecolor{C0}{HTML}{4C72B0}
\definecolor{C1}{HTML}{DD8452}
\definecolor{C2}{HTML}{55A868}
\definecolor{C3}{HTML}{C44E52}

\usepackage{gensymb}

\newcommand\beq{\begin{equation}}
\newcommand\eeq{\end{equation}}

\newcommand\btb{\begin{tcolorbox}}
\newcommand\etb{\end{tcolorbox}}

\begin{document}
	
	\title{Towards practical non-Markovianity measures: Normalization and regularization techniques}

    \author{L. A. Mazhorina}\email{l.akopyan@rqc.ru}
	\affiliation{Russian Quantum Center,  Skolkovo, Moscow 143025, Russia}

    \author{N. D. Korolev}
	\affiliation{Russian Quantum Center,  Skolkovo, Moscow 143025, Russia}

        \author{N. V. Morozov}
	\affiliation{Russian Quantum Center,  Skolkovo, Moscow 143025, Russia}

        \author{E. Yu. Egorova}
	\affiliation{Russian Quantum Center,  Skolkovo, Moscow 143025, Russia}
 \affiliation{National University of Science and Technology “MISIS”, Moscow 119049, Russia}

        \author{A. V. Zotova}
	\affiliation{Russian Quantum Center,  Skolkovo, Moscow 143025, Russia}
 \affiliation{National University of Science and Technology “MISIS”, Moscow 119049, Russia}
 \affiliation{Moscow Institute of Physics and Technology, Dolgoprudny 141701, Russia}

        \author{T. A. Chudakova}
	\affiliation{Russian Quantum Center,  Skolkovo, Moscow 143025, Russia}
 \affiliation{National University of Science and Technology “MISIS”, Moscow 119049, Russia}
 \affiliation{Moscow Institute of Physics and Technology, Dolgoprudny 141701, Russia}

 \author{G. S. Mazhorin}
	\affiliation{Russian Quantum Center,  Skolkovo, Moscow 143025, Russia}
    \affiliation{National University of Science and Technology “MISIS”, Moscow 119049, Russia}
    \affiliation{Moscow Institute of Physics and Technology, Dolgoprudny 141701, Russia}

          \author{N. Sterligov}
	\affiliation{Russian Quantum Center,  Skolkovo, Moscow 143025, Russia}

           \author{A. S. Kazmina}
	\affiliation{Russian Quantum Center,  Skolkovo, Moscow 143025, Russia}
 \affiliation{National University of Science and Technology “MISIS”, Moscow 119049, Russia}
 \affiliation{Moscow Institute of Physics and Technology, Dolgoprudny 141701, Russia}

         \author{A. M. Polyanskiy}
	\affiliation{Russian Quantum Center,  Skolkovo, Moscow 143025, Russia}
 \affiliation{National University of Science and Technology “MISIS”, Moscow 119049, Russia}
 \affiliation{Moscow Institute of Physics and Technology, Dolgoprudny 141701, Russia}

 \author{N. Abramov}
 \affiliation{National University of Science and Technology “MISIS”, Moscow 119049, Russia}

        \author{I. O. Gridnev}
	\affiliation{Russian Quantum Center,  Skolkovo, Moscow 143025, Russia}
    \affiliation{Moscow Institute of Physics and Technology, Dolgoprudny 141701, Russia}

   \author{M. A. Gavreev}
	\affiliation{Russian Quantum Center,  Skolkovo, Moscow 143025, Russia}
	
	\author{A. Matveev}
	\affiliation{Russian Quantum Center,  Skolkovo, Moscow 143025, Russia}

    	\author{O. Lakhmanskaya}
	\affiliation{Russian Quantum Center,  Skolkovo, Moscow 143025, Russia}
    
    \author{I. A. Simakov}
	\affiliation{Russian Quantum Center,  Skolkovo, Moscow 143025, Russia}
 \affiliation{National University of Science and Technology “MISIS”, Moscow 119049, Russia}
 \affiliation{Moscow Institute of Physics and Technology, Dolgoprudny 141701, Russia}

 	\author{E. A. Polyakov}
        \affiliation{Russian Quantum Center,  Skolkovo, Moscow 143025, Russia}

        \author{K. Lakhmanskiy}
	\affiliation{Russian Quantum Center,  Skolkovo, Moscow 143025, Russia}

	\date{\today}
	
\begin{abstract}

Measures characterizing the non-Markovianity degree of the quantum dynamics have several drawbacks when applied to real devices. They depend on the chosen measurement time interval and are highly sensitive to experimental noise and errors. We propose several techniques to enhance the practical applicability of the measures and verify our findings experimentally on a superconducting transmon and a trapped-ion qubit. The time dependence can be disregarded by introduction of the measure per oscillation, while the sensitivity to noise is reduced by applying a regularization procedure. The results for both types of qubits are compared with theoretical predictions for a simple model of non-Markovianity based on qubit-qubit interaction.

\end{abstract}

	\maketitle
	
			

Noise limits the performance of state-of-the-art quantum processors. Using the formalism of open quantum systems, errors can be treated as auxiliary degrees of freedom interacting with the system, collectively called the environment. The reduced dynamics of the open quantum system can be described by a master equation in the Lindblad form \cite{Lindblad76a, AgarwalBook, FicekBook, NielsenChuang2010, breuer2007}, assuming the absence of memory effects in the system dynamics (Markov approximation). 

Leakage out of the computational subspace, fluctuations in driving fields and qubit parameters, sample defects, and other error sources induce quantum dynamics that violate the Markov approximation \cite{hashim2024, Wallman2016, Lisenfeld2015, krantz2019, Gulacsi2023, agarwal2023, Nakamura2024, Nakamura2024a1, velazquez2025}. This behavior, known as non-Markovian, requires specialized methods for error correction, description, and characterization \cite{terhal2005, White2020, Pedro2021, McEwen2021, white2023}. 

During the past two decades, a variety of approaches have been developed to characterize the degree of non-Markovianity \cite{wolf2008, Breuer2009, rivas2010, Laine2010, Luo2012, bylicka2014, Guo2022}. Measures based on the revival of correlations and entanglement between the system and an ancilla \cite{Luo2012, rivas2010}, and distinguishability quantifiers \cite{Breuer2009, Megier2021} have been theoretically introduced, compared with each other \cite{Mannone2013, Addis2014, settimo2022}, and experimentally demonstrated (see Refs. \cite{Liu2011, Gessner2014, Li2022, Gaikwad2024} and references therein). However, these measures have two key drawbacks. First, they are highly sensitive to noise in the experimental data. Second, the values of the measures are influenced by the Markovian properties of the dynamics and depend on the measurement time interval. These issues make it challenging to experimentally determine the degree of non-Markovianity and to compare the non-Markovian properties of different experimental devices. A truly useful measure should be an experimentally accessible, dimensionless quantity that makes it possible to distinguish the non-Markovianity of the dynamics from Markovianity and noise. It must have a clear physical interpretation, be robust against experimental noise, and remain measurable even when the characteristic timescales of memory effects are unknown.

In this work, we address these challenges and consider the Breuer, Laine, and Piilo (BLP) trace distance-based measure \cite{Breuer2009}. We propose a regularization method that reduces the influence of experimental noise on the measured degree of non-Markovianity. Our approach introduces criteria that filter out experimental noise while preserving genuine non-Markovian effects. Furthermore, we overcome the dependence of the measure on the timescale by introducing the BLP measure per oscillation, which is normalized and independent of the chosen time interval. The measure per oscillation has a clear physical interpretation: It shows which fraction of information is coherently exchanged between the system and the environment. We experimentally demonstrate the proposed ideas using two different quantum systems: a superconducting transmon and a trapped-ion qubit. Our approach enables the experimental determination of the measure for both systems, allowing for a direct comparison even though their dynamical timescales differ by three orders of magnitude. We successfully validated our regularization method using a well-defined model, achieving strong agreement between the regularized experimental and theoretical values of the measure. 

The BLP measure \cite{Breuer2009, Laine2010} is based on the calculation of the trace distance between two initial states of the system $\rho_1(0)$ and $\rho_2(0)$ evolving in time:
\beq D(\rho_1(t), \rho_2(t)) = \frac{1}{2}||\rho_1(t) - \rho_2(t)||, \quad ||A|| = \Tr\sqrt{A^{\dagger}A}. \eeq
For Markovian dynamics, the trace distance is a monotonically decreasing function of time for all pairs of initial states with a negative derivative $\sigma_t(\rho_1, \rho_2) = \dfrac{d}{dt} D(\rho_1(t), \rho_2(t)) <0$ \cite{Breuer2009, Addis2014, Nakamura2024}. Information for such a system only dissipates to the environment without a backflow. When the dynamics is non-Markovian, the trace distance becomes a non-monotonic function and increases with time, showing a revival of information from the environment back to the system during some time intervals. The measure accounts for all events of non-Markovian backflow of information to the system:
\beq\label{eq:BLPchimax}
\chi = \max\limits_{\rho_1, \rho_2} \int\limits_{0}^{+\infty} dt\,\sigma_t(\rho_1, \rho_2) \bigg |_{\sigma_t > 0}.
\eeq
The maximum in Eq. \eqref{eq:BLPchimax} is taken over all pairs of initial states. 

We note that the calculation of the BLP measure implies integration over the infinite time interval, while in experiments the time step and the measurement time are finite. In this case Eq. \eqref{eq:BLPchimax} reduces to the sum of the positive differences $D(m+1)-D(m)>0$ for all the measured data points at neighboring time steps $m$ and $m+1$ \cite{Breuer2009}. Hence, the measure depends on the measurement time. Moreover, if the measured trace distances are affected by experimental errors, then false positive or negative differences $D(m+1)-D(m)$ between some neighboring points can occur and lead to fictitious growth or decrease of the BLP measure. This means that the BLP measure \eqref{eq:BLPchimax} is ill-conditioned: It is unstable with respect to the experimental fluctuations and can not accurately distinguish them from the true non-Markovianity of the dynamics.

We perform experiments on two types of physical qubits: a superconducting transmon and a trapped-ion qubit. Both qubits have long coherence times compared to the duration of operations, and relaxation of quantum states does not significantly affect the measured trajectories and trace distances. The dynamics of each qubit is described by a simple model of non-Markovianity in which the reservoir degrees of freedom are represented as another qubit. The model allows us to make analytical predictions for the BLP measure, to which we compare the experimentally obtained values. The model Hamiltonian has the following form:
\beq\label{eq::HamSpinQub}
    H = \hbar \, \dfrac{\Omega}{2} \, \sigma_z + \hbar \, \dfrac{\omega}{2} \, \sigma_z^R + \hbar \,  \dfrac{g}{2} \left(\sigma_- \sigma_+^R + \sigma_+ \sigma_-^R \right),
\eeq
where the first term is the system qubit Hamiltonian with transition frequency $\Omega$, $\sigma_{\pm, z}$ are the raising/lowering and Pauli-z qubit operators, the superscript $R$ refers to the reservoir qubit, $\omega$ is the frequency of the reservoir qubit, and $g$ is the coupling strength between two qubits. The qubits are isolated and evolve under unitary evolution with Hamiltonian \eqref{eq::HamSpinQub} from some initial separable state. The initial state of the reservoir is the ground state $\ket{0}$. For the reduced density matrix of a system qubit, the maximum in Eq. \eqref{eq:BLPchimax} is reached for the ground $\ket{0}$ and excited $\ket{1}$ initial states, and the trace distance for this optimal pair of states has the following dependence on time (see Supplemental Material \cite{supp} and Ref. \cite{Wissmann2012} for the derivation of this equation and the definition of the optimal pair of states, respectively):
\beq\label{eq::transmonTD}
    D_{\textrm{opt}}(t) = 1 - \frac{g^2 \sin^2{\left (t\dfrac{\sqrt{g^2 + \Delta^2}}{2}\right )}}{g^2 + \Delta^2}.
\eeq
Here, $\Delta = \Omega-\omega$ is the detuning of the system qubit frequency from the reservoir qubit frequency. The optimal trace distance \eqref{eq::transmonTD} oscillates in time at a single frequency $\sqrt{g^2 + \Delta^2}$. For systems with this property, it is possible to introduce the BLP measure per oscillation $\chi^{2\pi}$, whose value is calculated as the difference between the maximum and the minimum of $D_{opt}(t)$ in \eqref{eq::transmonTD}:
\beq\label{eq:ChiperOsci}
\chi^{2\pi} = \frac{g^2}{g^2 + \Delta^2}.
\eeq

The BLP measure per oscillation \eqref{eq:ChiperOsci} has several attractive properties. First, its value ranges from 0 to 1 and is independent of the measurement time and type of the system. This makes it suitable for the comparison of the non-Markovian properties of different systems. Second, Eq. \eqref{eq:ChiperOsci} suggests a recipe on how to create and control the degree of non-Markovianity for the system by setting the non-zero system-reservoir coupling constant $g\neq 0$ and the system qubit frequency near the reservoir qubit frequency $ \Omega \approx \omega$ with maximum degree of non-Markovianity $\chi^{2\pi} = 1$ at resonance $\Omega = \omega$. The BLP measure per oscillation is the portion of information that is coherently exchanged between the system and the environment (see also Ref. \cite{Breuer2009}). Environments with high $\chi^{2\pi}$ can become beneficial non-Markovian quantum memories \cite{bylicka2014, Man2015}.

For each qubit, we measure the time dependence of the reduced system density matrix for different initial states (see examples of trajectories in the Supplemental Material \cite{supp}). For measured trajectories we find the trace distances and calculate the state-pair-dependent BLP measure per oscillation $\chi^{2\pi}_{ij}$ between states $\rho_i$ and $\rho_j$:
\beq\label{eq:chiij}
\chi^{2\pi}_{ij} =\int dt \, \sigma_t(\rho_i, \rho_j) \bigg |_{\sigma_t > 0}.
\eeq 
The integral in Eq. \eqref{eq:chiij} can be calculated over a single period up to time $\dfrac{2\pi}{\sqrt{g^2 + \Delta^2}}$ or by averaging over several periods for the entire measured time interval. In this work, we used the latter method. 

\begin{figure}
    \centering
    \includegraphics[width=1\linewidth]{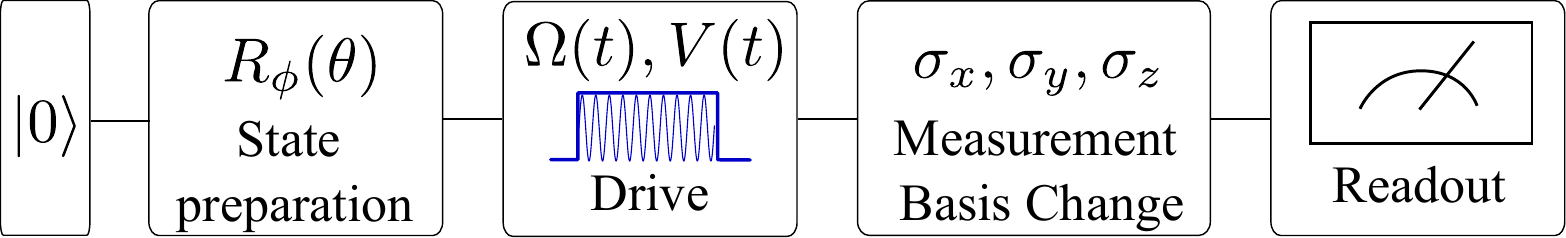}
    \caption{Pulse sequence for non-Markovianity measurement on a single qubit. For a trapped-ion qubit the drive pulse is the resonant laser pulse which Rabi frequency is equal to the motional frequency, while for a transmon qubit the drive is the voltage, acting on a coupler.}
    \label{fig:PulseSeq}
\end{figure}

Measurements on a transmon qubit were performed to test the analytical expressions obtained for the model and include the calculation of the BLP measure per oscillation \eqref{eq:ChiperOsci} on the optimal pair, while for a trapped-ion qubit we also analyze the influence of noise occurring on the state-pair-dependent BLP measure per oscillation \eqref{eq:chiij} and demonstrate the idea of measure regularization \cite{supp}. The pulse sequence used for the measurements is shown in Fig. \ref{fig:PulseSeq}. The non-Markovian system-reservoir coupling drive field with variable duration controls the frequency detuning $\Delta$ and the coupling strength $g,$ and hence the degree of non-Markovianity of the system quantum dynamics. It is turned off during all other steps.

For non-Markovianity measurements on a transmon qubit, we use the transmon quantum processor described in Ref. \cite{egorova2024}, where a tunable coupler is used to control the interaction between two qubits and perform quantum gates. We consider one of the qubits as the system, and the other qubit as the corresponding environment. The Hamiltonian of interaction between two qubits is given exactly with Eq. \eqref{eq::HamSpinQub}, where $\Omega,\omega$ are the frequencies of the system and the reservoir qubits, respectively, $g$ is the XX-interaction coupling strength between the qubits in the rotating wave approximation. The coupling strength $g$ between two qubits and the difference between their frequencies $\Delta$ are simultaneously adjusted by external flux applied to the coupler. By changing the coupler flux with DC voltage, one can control the degree of non-Markovianity mediated by the interaction between two qubits.

\begin{figure*}
    \centering
    \includegraphics[width=1\linewidth]{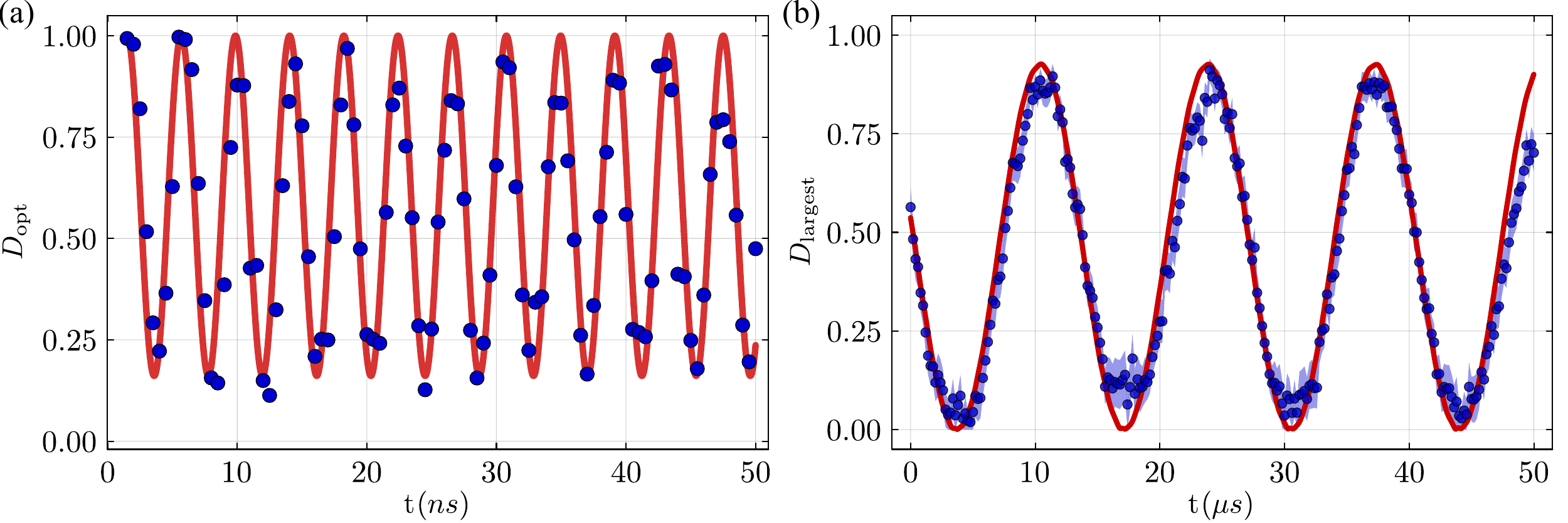}
    \caption{Experimental trace distance depending on time for optimal pair of initial states $\ket{0}$ and $\ket{1}$ of the transmon qubit (a) and for the pair with the largest experimentally obtained measure of the trapped-ion qubit (b). Blue dots represent experimental data, while the solid red lines depict the results of a simulation with Hamiltonian \eqref{eq::HamSpinQub}. One standard deviation of $500$ experimental shots of the ion qubit is shown with a blue region in (b). See Supplemental Material \cite{supp} for the details on error calculation.}
    \label{fig:TransmonNM}
\end{figure*}

We fit the experimentally obtained quantum trajectories with the model Hamiltonian \eqref{eq::HamSpinQub} and find the following parameters for the transmon qubit in the rotating frame:
$ \omega = -2\pi\times37 $ MHz, $\Omega = 2\pi\times59$ MHz, $ g = 2\pi\times219 $ MHz. The trace distance for the optimal pair of initial states $\ket{0}$ and $\ket{1}$ is shown in Fig. \ref{fig:TransmonNM}(a). It starts from $1$ and does not reach the $0$ value because the detuning $\Delta \neq 0$. The trace distance oscillates at frequency $2\pi\times24$ MHz, while the BLP measure per oscillation is equal to $\chi^{2\pi}_{\textrm{exp}}=0.74$. This value differs from the theoretically derived measure for these parameters $\chi^{2\pi}_{\textrm{model}}=0.84$. We associate this discrepancy with a large time step used in the measurements, which was limited to $0.5$ ns due to technical reasons, and experimental fluctuations.

For a trapped ion non-Markovianity measurements, we used a single $^{40}Ca^+$ ion in a linear Paul trap. The ion stored in a trap has several degrees of freedom, including the internal electronic levels and the motion of its center of mass. Two of the electronic levels connected by the optical quadrupole transition are used as a qubit. The control over qubit degrees of freedom is realized via a classical laser acting at resonance with the qubit transition. Proper trapping and cooling of the ion to the motional ground state allows us to quantize the center of mass motion and consider it as another qubit with frequency $\omega,$ controlled by the voltages set to the trap electrodes. The laser shining on the ion mediates the interaction between the qubit and the motional degrees of freedom. For non-Markovianity measurements, we consider the qubit degree of freedom as the open quantum system and the motion as the corresponding environment (see also Refs. \cite{Li2022, Gessner2014}). Description of the experimental setup and the full laser-ion Hamiltonian can be found in the Supplemental material \cite{supp} and Refs. \cite{Pogorelov2021, DD2003, Akopyan2022}.

The Hamiltonian of laser-ion interaction after expansion to the first order in the Lamb-Dicke parameter $\eta$, a unitary transformation, and in a rotating wave approximation has the form of Eq. \eqref{eq::HamSpinQub} with $\Omega$ being the Rabi frequency of the laser-ion interaction and the coupling strength $g = \Omega \,\eta.$ The Rabi frequency $\Omega$ that determines the system qubit frequency and the system-reservoir coupling $g$ is controlled by a laser power. For the measurements, we created the maximal degree of non-Markovianity setting the Rabi frequency of the laser-ion interaction at resonance with the motional frequency $\Omega = \omega = 2\pi \times 591$ kHz and prepared four arbitrary initial states. The qubit-mode coupling constant $g$ and hence the frequency of trace distance oscillations \eqref{eq::transmonTD} are equal to $2\pi\times74$ kHz. Due to limitations of the control electronics, the minimal pulse duration in the measurements was $3$ $\mu$s, and the initial time shown in all figures is shifted by this duration. The trace distance for a pair of states with the largest measured value of the BLP state-pair-dependent measure per oscillation is shown in Fig. \ref{fig:TransmonNM}(b). The blue dots represent experimental data, while the solid red lines depict the numerical simulation of the experiment with Hamiltonian \eqref{eq::HamSpinQub} with ideal parameters that were set to the experiment. Because of zero detuning $\Delta = 0$, the trace distance reaches $0$ indicating the full transfer of information from the system to the environment, but it does not become $1$ because the pair of initial states is not orthogonal.

\begin{figure*}[t]
     \centering
     \includegraphics[width=1.0\linewidth]{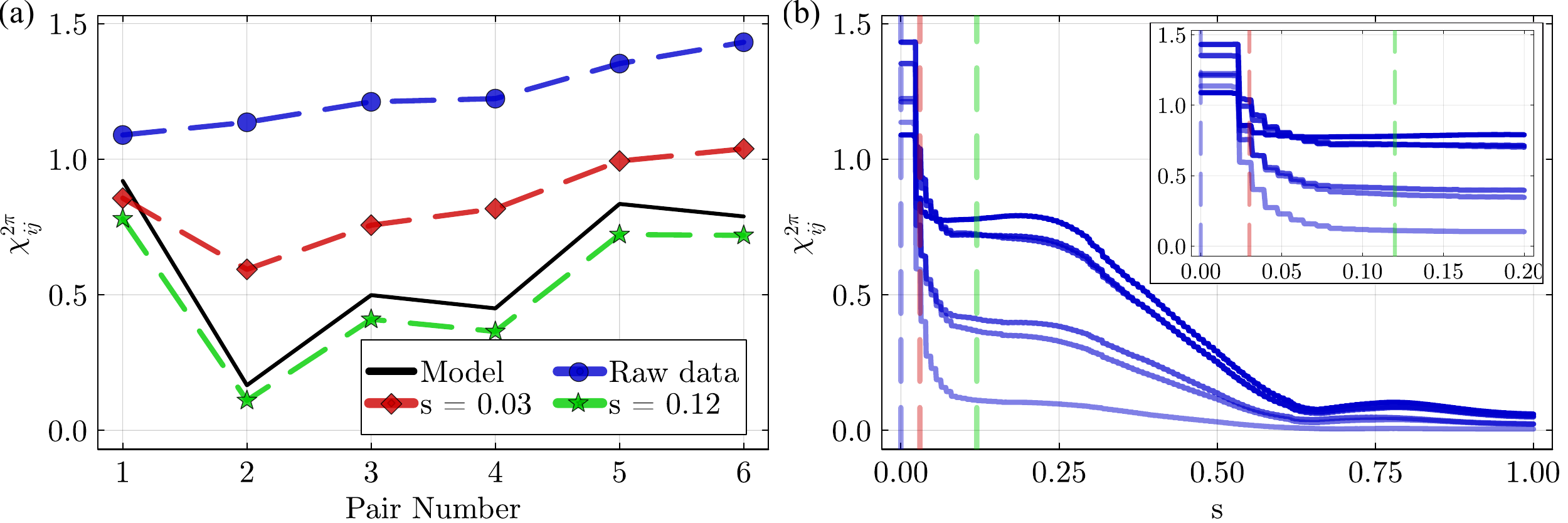}
     \caption{Regularized state-pair-dependent non-Markovianity BLP measure per oscillation for measured pairs of initial states of the ion qubit depending on (a) the ion initial state pair number and (b) the smoothing parameter of the algorithm. (a) The solid black line is the simulation with ideal parameters set to the experiment. Colored lines with markers represent the measure calculated for experimental data with different degrees of smoothing. The degree of smoothing $s$ used in Loess algorithm is shown on legend. The blue circles are the measures for raw data. 
     (b) Solid lines represent the regularized experimentally obtained measure for different initial state pairs. Each curve contains initial noise peak exceeding theoretical values of the measure and reaches plateau. The plateau shows the range of the smoothing parameters for which the experimental  noise is successfully eliminated. Three vertical dashed lines show the degrees of smoothing $s$ which are used in plot (a).}
     \label{fig:IonChiOnS}
\end{figure*}

Figure \ref{fig:IonChiOnS}(a) shows the values of the experimental and theoretical BLP state-pair-dependent measure per oscillation for the 6 measured initial state pairs. The points are connected by lines for clarity. The solid black line without markers shows the measures for simulated data, while the dashed blue line with circles represents the experimentally obtained measures. The non-Markovianity measures for raw experimental data significantly exceed the simulated ones because of experimental noise and fluctuations.

To decrease the influence of noise, we suggest to smooth the experimentally obtained trace distances and calculate the non-Markovianity measure on smoothed trace distances. To demonstrate the idea, we have chosen the Loess algorithm \cite{Cleveland1979}, in which the smoothing is done by fitting the neighborhood of each point with degree 2 polynomials. The real number $s \in [0, 1]$ defines the smoothing degree, which is proportional to the number of points in the neighborhood that is taken for the fitting procedure. The larger the smoothing degree, the smoother the resulting curve. Examples of the measured trace distances and the influence of smoothing on them are given in Supplemental material \cite{supp}. 
To make a quantitative comparison of smoothed and theoretical curves and provide the algorithm on how to choose the smoothing degree, we calculated the state-pair-dependent BLP measure per oscillation for experimentally measured data for different degrees of smoothing. The results for all pairs of measured trajectories are shown in Fig. \ref{fig:IonChiOnS}(b) and contain a horizontal plateau, at which the measure does not change during the same interval of the smoothing degree. We associate these plateaus with the smoothing degree region, at which the experimental noise is removed, while the true non-Markovian oscillations are not affected by a smoothing procedure. Further increase of the degree of smoothing starts to eliminate these oscillations, and at the maximum value of the smoothing degree $s = 1$ leads to an almost zero measure. A numerical verification of this behavior with simulated noise can be found in the Supplemental material \cite{supp}.

This result suggests the algorithm on how to choose the smoothing degree, at which the noise contribution to the BLP measure is eliminated. One should take a smoothing degree, at which the values of the measure calculated on a smoothed trace distance lie on the plateau. The measures, regularized with this rule, are illustrated in Fig. \ref{fig:IonChiOnS}(a) with star markers. The diamond markers in Fig. \ref{fig:IonChiOnS}(a) are calculated for the degree of smoothing at which the plateau on the measures has not been reached yet. The corresponding degrees of smoothing are shown in Fig. \ref{fig:IonChiOnS}(b) with dashed vertical lines. The regularized BLP state-pair-dependent measure presented in Fig. \ref{fig:IonChiOnS}(a) with stars reproduces the simulated results with $66 - 91\, \%$ of coincidence, while the measures without regularization show non-physical values exceeding $1$.

\begin{figure}[tb]
\centering
\includegraphics[width=0.5\textwidth]{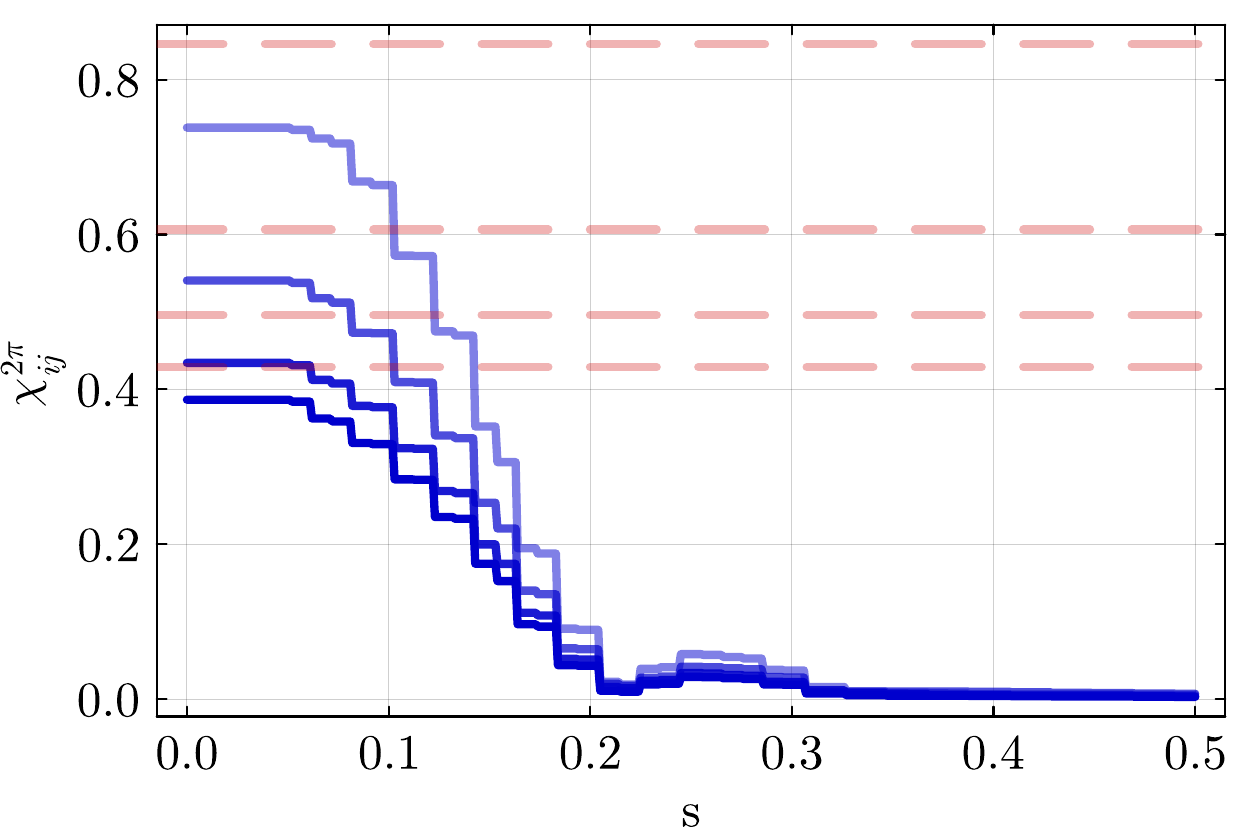}
\caption{The state-pair-dependent BLP measure per oscillation as a function of the smoothing degree for the transmon qubit for several initial state pairs. Dashed horizontal lines show the theoretical values of the measure.}
\label{fig:transmonChiOnS}
\end{figure}

The suggested regularization technique can be used when the oscillations of the trace distance are measured with enough resolution in time. As shown in Fig. \ref{fig:IonChiOnS}(b), in this case, the dependence of the measure on time contains the noise peak for small smoothing degrees and the plateau on which the noise is eliminated. Contrary, in measurements with large time steps the noise peak and, as a consequence, the left boundary of the plateau are absent. This means that experimental errors cannot be removed by smoothing. The illustration of this case is given in Fig. \ref{fig:transmonChiOnS} for the transmon qubit. Fitting of trace distances with analytical curves \eqref{eq::transmonTD} can be used in this case if the behavior of the trace distance is known.

The methods described in this work are applicable to most existing non-Markovianity measures \cite{Addis2014}. The corresponding measure per oscillation is defined for non-decaying systems that oscillate at a single dominant frequency. Generalizations of the measures per oscillation have to be done for the cases of decaying systems and systems with trace distances oscillating at several frequencies.

In this work, we have not analyzed the influence of systematic experimental errors on non-Markovianity measure. In addition, other methods of signal filtration can be considered to clear the data from the occurring noise. To name a few, Fourier filtration can be useful in cases where the range of physical frequencies on which the oscillations of the system occur is known, and data-driven techniques \cite{Luchnikov2020, Luchnikov2022} for analysis, filtration, and prediction of non-Markovian quantum dynamics can be used. These questions are a topic for future research.

In conclusion, we experimentally realized the non-Markovian quantum dynamics on a superconducting transmon and a trapped-ion qubit and quantitatively characterized the degree of non-Markovianity for each qubit. We proposed and experimentally calculated for the transmon qubit the BLP measure per oscillation that characterizes the degree to which information is coherently exchanged between the system and the reservoir. This type of measure allows for the comparison of different types of physical systems with respect to their non-Markovian properties. For the trapped-ion qubit, we measured the state-pair-dependent BLP measure per oscillation and showed that it is sensitive to experimental fluctuations and perceives experimental noise as non-Markovianity. To stabilize the measure and remove noise-induced false non-Markovianity, we introduced a technique of the BLP measure regularization based on the smoothing of corresponding experimental trace distances and verified it experimentally. With this regularization, we significantly improved the agreement between the simulated and experimental non-Markovianity measures for the trapped-ion qubit for all measured initial state pairs.

This work was supported by Rosatom in the framework of the Roadmap for Quantum computing (Contract No. 868-1.3-15/15-2021 dated October 5 and No. 151/21-503, 2021 dated 12/21/2021). We acknowledge partial support of the Ministry of Science and Higher Education of the Russian Federation in the framework of the Program of Strategic Academic Leadership “Priority 2030” (Strategic Project Quantum Internet).
The authors are grateful to Alexey Ustinov for critical comments on the manuscript. 
We gratefully acknowledge A. Prozorovskaya for the objective design for the trapped-ion global addressing setup. 
We express our gratitude to I.S. Besedin for his significant contribution to the design of the superconducting device. 
We also thank D.A. Kalacheva, V.B. Lubsanov and A.N. Bolgar for fabricating the sample.

	\nolinenumbers
	\bibliography{lib.bib}
	
	\nolinenumbers

\appendix
 
\end{document}


\title{Supplemental material for Towards practical non-Markovianity measures: Normalization and regularization techniques}

    \author{L. A. Mazhorina}\email{l.akopyan@rqc.ru}
	\affiliation{Russian Quantum Center,  Skolkovo, Moscow 143025, Russia}

    \author{N. D. Korolev}
	\affiliation{Russian Quantum Center,  Skolkovo, Moscow 143025, Russia}

        \author{N. V. Morozov}
	\affiliation{Russian Quantum Center,  Skolkovo, Moscow 143025, Russia}

        \author{E. Yu. Egorova}
	\affiliation{Russian Quantum Center,  Skolkovo, Moscow 143025, Russia}
 \affiliation{National University of Science and Technology “MISIS”, Moscow 119049, Russia}

        \author{A. V. Zotova}
	\affiliation{Russian Quantum Center,  Skolkovo, Moscow 143025, Russia}
 \affiliation{National University of Science and Technology “MISIS”, Moscow 119049, Russia}
 \affiliation{Moscow Institute of Physics and Technology, Dolgoprudny 141701, Russia}

        \author{T. A. Chudakova}
	\affiliation{Russian Quantum Center,  Skolkovo, Moscow 143025, Russia}
 \affiliation{National University of Science and Technology “MISIS”, Moscow 119049, Russia}
 \affiliation{Moscow Institute of Physics and Technology, Dolgoprudny 141701, Russia}

 \author{G. S. Mazhorin}
	\affiliation{Russian Quantum Center,  Skolkovo, Moscow 143025, Russia}
    \affiliation{National University of Science and Technology “MISIS”, Moscow 119049, Russia}
    \affiliation{Moscow Institute of Physics and Technology, Dolgoprudny 141701, Russia}

          \author{N. Sterligov}
	\affiliation{Russian Quantum Center,  Skolkovo, Moscow 143025, Russia}

           \author{A. S. Kazmina}
	\affiliation{Russian Quantum Center,  Skolkovo, Moscow 143025, Russia}
 \affiliation{National University of Science and Technology “MISIS”, Moscow 119049, Russia}
 \affiliation{Moscow Institute of Physics and Technology, Dolgoprudny 141701, Russia}

         \author{A. M. Polyanskiy}
	\affiliation{Russian Quantum Center,  Skolkovo, Moscow 143025, Russia}
 \affiliation{National University of Science and Technology “MISIS”, Moscow 119049, Russia}
 \affiliation{Moscow Institute of Physics and Technology, Dolgoprudny 141701, Russia}

 \author{N. Abramov}
 \affiliation{National University of Science and Technology “MISIS”, Moscow 119049, Russia}

        \author{I. O. Gridnev}
	\affiliation{Russian Quantum Center,  Skolkovo, Moscow 143025, Russia}
    \affiliation{Moscow Institute of Physics and Technology, Dolgoprudny 141701, Russia}

   \author{M. A. Gavreev}
	\affiliation{Russian Quantum Center,  Skolkovo, Moscow 143025, Russia}
	
	\author{A. Matveev}
	\affiliation{Russian Quantum Center,  Skolkovo, Moscow 143025, Russia}

    	\author{O. Lakhmanskaya}
	\affiliation{Russian Quantum Center,  Skolkovo, Moscow 143025, Russia}
    
    \author{I. A. Simakov}
	\affiliation{Russian Quantum Center,  Skolkovo, Moscow 143025, Russia}
 \affiliation{National University of Science and Technology “MISIS”, Moscow 119049, Russia}
 \affiliation{Moscow Institute of Physics and Technology, Dolgoprudny 141701, Russia}

 	\author{E. A. Polyakov}
        \affiliation{Russian Quantum Center,  Skolkovo, Moscow 143025, Russia}

        \author{K. Lakhmanskiy}
	\affiliation{Russian Quantum Center,  Skolkovo, Moscow 143025, Russia}

	\date{\today}

	\maketitle

\onecolumngrid

\section{Derivation of trace distance for considered model}\label{App::AnalyticsTDModel}

The model Hamiltonian (see Eq. (3) of the main  manuscript) has the following eigenfrequencies (here we set $\hbar = 1$):
\beq\label{eq:Eigenvals}
    \begin{cases}
        2\lambda_{1} = \omega + \Omega,\\
        2\lambda_{2} = -\sqrt{g^{2}+\Delta^{2}},\\
        2\lambda_{3} = \sqrt{g^{2}+\Delta^{2}},\\
        2\lambda_{4} = -\omega - \Omega.\\
    \end{cases}
\eeq
The matrix diagonalizing the Hamiltonian reads:
\beq
    S =\left(
    \begin{array}{cccc}
    1 & 0 & 0 & 0 \\
    0 & -\frac{\sqrt{g^{2}+\Delta^{2}} +\Delta}{g}& \frac{\sqrt{g^{2}+\Delta^{2}} - \Delta}{g} & 0 \\
    0 & 1 & 1 & 0 \\
    0 & 0 & 0 & 1 \\
    \end{array}
    \right ).
\eeq

Using these expressions, one can find the system-reservoir evolution operator:
\beq
    U(t) = \exp(- \frac{i H t}{\hbar}) = S \exp(- i \Lambda t) S^{-1},
\eeq
where $\Lambda$ is the diagonal matrix consisting of eigenvalues \ref{eq:Eigenvals}.
The system qubit state is the partial trace over reservoir of the full density matrix:

\beq
    \rho(t) =  \text{Tr}_R \left [U(t) \ket{\psi(0)0}\bra{\psi(0)0} U^\dagger(t) \right ],
\eeq
where $\ket{\psi(0)}$ is some initial pure state of the system qubit, and the initial state of the reservoir qubit is fixed to $\ket{0}$.

The optimal pair of initial states of a qubit for which the BLP measure (see Eq. (2) of the main manuscript) is calculated is some pair of orthogonal initial states [36]. An arbitrary pair of orthogonal states $\ket{\psi_0}$ and $\ket{\psi_1}$ of a system qubit has the following general form:
\beq
\begin{split}
    &\ket{\psi_0} = A \ket{0} + B e^{i\beta}\ket{1},\\
    &\ket{\psi_1} = B  \ket{0} - A e^{i\beta}\ket{1},
\end{split}
\eeq
where $A$ and $B$ are real numbers with $A^2 + B^2 = 1$, and $\beta$ is the phase of the state.

Trace distance for this pair reads:
\beq
    D_{12} (t) = \sqrt{\frac{\left(1-2A^{2}\right)^{2}\left(\Delta^{2}+g^{2}\cos^{2}\left(t\dfrac{\sqrt{g^{2}+\Delta^{2}}}{2}\right)\right)^{2}}{\left(g^{2}+\Delta^{2}\right)^{2}}+\frac{4A^{2}\left(1-A^{2}\right)\left(g^{2}\cos^{2}\left(t\dfrac{\sqrt{g^{2}+\Delta^{2}}}{2}\right)+\Delta^{2}\right)}{g^{2}+\Delta^{2}}}
\eeq

The BLP measure per oscillation for this pair of initial states then becomes:

\beq\label{eq:TDArbOrtPair}
    \chi_{12}^{2\pi} = D_{12} (t_{max}) - D_{12} (t_{min}) =
    1 - \sqrt{\frac{\left(1-2A^{2}\right)^{2}\Delta^{4}}{\left(g^{2}+\Delta^{2}\right)^{2}}+\frac{4A^{2}\left(1-A^{2}\right)\Delta^{2}}{g^{2}+\Delta^{2}}}.
\eeq
One can see that the optimal pair of orthogonal initial states is the pair of states $\ket{0}$ and $\ket{1}$ with $A=0$ and obtain the formula (5) of the main manuscript.

\section{Measurement details and statistical errors}\label{app:Measdet}

The quantum state of a single qubit can be described in the Pauli basis $\hat{\vec{\sigma}}$:
\beq\label{eq:SQRho}
    \hat{\rho} = \frac{\mathcal{I}+\vec{v}\hat{\vec{\sigma}}}{2},
\eeq
where $\mathcal{I}$ is the identity operator and the components of Bloch vector $v_k, \, k \in (x, y, z)$ can be obtained with the measurement of the state in the corresponding Pauli basis $\sigma_k$. Using this decomposition, one can calculate the trace distance $D_{ij}(t)$ for an arbitrary evolving pair of qubit initial states $\rho_i(0)$ and $\rho_j(0)$ on time:
\beq\label{eq:TDDecomp}
    D_{ij} (t) = \frac{1}{2} Tr\sqrt{(\rho_i (t) - \rho_j (t))^2} = \dfrac{|\vec{v}_i(t)-\vec{v}_j(t)|}{2}.
\eeq
This relation holds for both a pure and a mixed qubit states. 

The description of the pulse sequence steps for the density matrix measurements shown in Fig. 1 of the main manuscript is the following.
\begin{enumerate}
    \item Initialization of the qubit to the $\ket{0}$ state;
    \item preparation of initial state;
    \item non-Markovian pulse for variable duration;
    \item change of the measurement basis;
    \item measurement of the final state in the chosen basis.
\end{enumerate}

The state preparation and change of the measurement basis are usually done with the same type of single-qubit unitary operations:
\begin{equation}\label{eq:idealRgate}
    R_{\phi}(\theta) = \left(
\begin{array}{cc}
\cos\,(\frac{\theta}{2}) & -i e^{-i\phi}\sin\, (\frac{\theta}{2}) \\
\\
-i e^{i\phi}\sin\, (\frac{\theta}{2}) & \cos\, (\frac{\theta}{2})
\end{array}
\right),
\end{equation}
where $\phi$ is the phase of the driving field acting on qubit at resonance with the qubit transition frequency, angle $\theta=\tau\Omega_{\textrm{Rabi}}$ is controlled by the duration of the pulse $\tau$ and Rabi frequency $\Omega_{\textrm{Rabi}}$ which defines the coupling strength between the qubit and the control field. 

This experimental sequence allows us to measure the components of Bloch vector $\vec{v}$ under non-Markovian evolution. If the norm of the measured Bloch vector exceeds 1 due to statistical noise, the renormalization $\vec{v} \rightarrow \dfrac{\vec{v}}{|\vec{v}|}$ follows the measurement. After the calculation of the Bloch vector, the trace distances between all pairs of measured states are calculated with Eq. \eqref{eq:TDDecomp}. The BLP state-pair-dependent measure is calculated as the sum of all positive differences of neighboring values of the trace distance, as described in the main text. The corresponding measure per oscillation is calculated as the average over all measured oscillations.

In the transmon measurements, we prepare six different initial states corresponding to the eigenstates of Pauli operators $\sigma_x, \sigma_y,  \sigma_z$, and apply a voltage of $0.6$ V to the coupler,  which approximately corresponds to a half flux quantum point. 

The Rabi frequency $\Omega_{\textrm{Rabi}}$ of a laser-ion interaction in the trapped-ion experiment was set equal to $2\pi\times187.5$ kHz, which is small compared to the motional frequency $\omega=2\pi\times591$ kHz. The experimental parameters of the prepared initial states for the trapped-ion qubit are shown in Table \ref{tab:IonStatesParameters}.

\begin{figure*}[tb]
\centering
\includegraphics[width=1.0\textwidth]{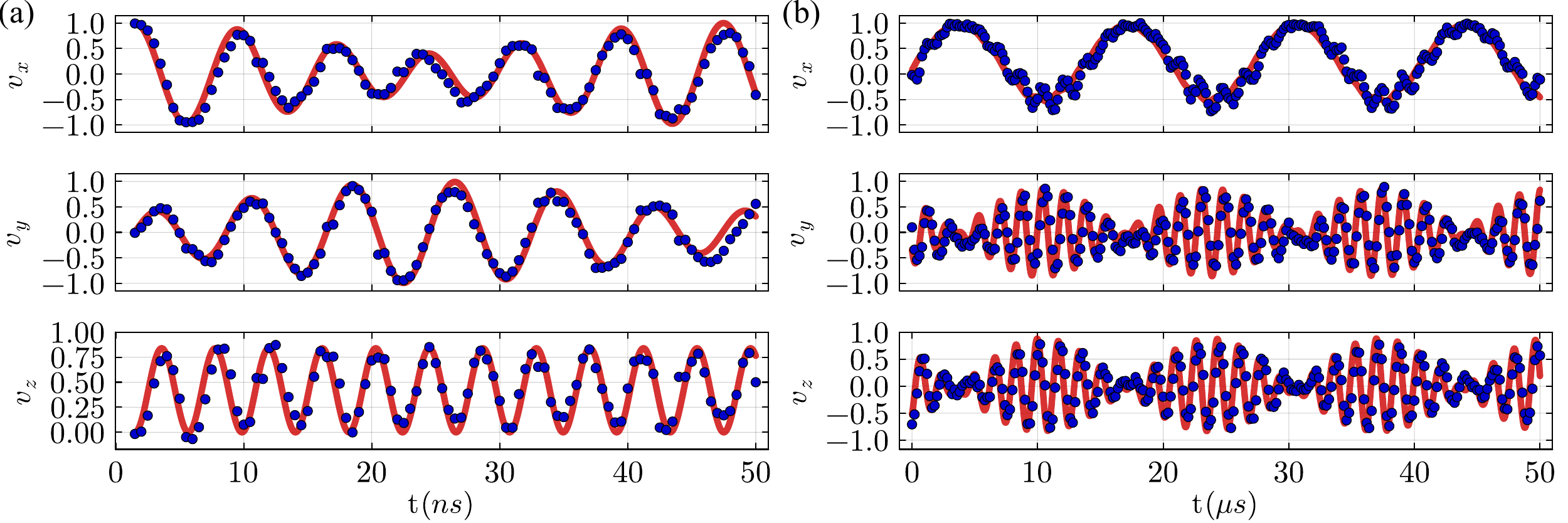}
\caption{Examples of trajectories for the transmon (a) and the trapped-ion (b) qubits. For the transmon qubit the state $\dfrac{\ket{0} + \ket{1}}{\sqrt{2}}$ is shown, while for the trapped-ion qubit the state $2$ from Table \ref{tab:IonStatesParameters} is presented.}
\label{fig:trajExamples}
\end{figure*}

The components of the Bloch vector $v_i$ for the state $\dfrac{\ket{0} + \ket{1}}{\sqrt{2}}$ of the transmon qubit and for the state (2) of the trapped-ion qubit are shown in Fig. \ref{fig:trajExamples}. The blue dots represent experimental data, while the solid red lines depict the numerical simulation of the experiment with Hamiltonian (3) given in the main manuscript with fitted data for the transmon qubit (a) and with ideal parameters that were set to the experiment for the trapped-ion qubit (b).

\begin{figure}
    \centering
    \includegraphics[width=0.5\linewidth]{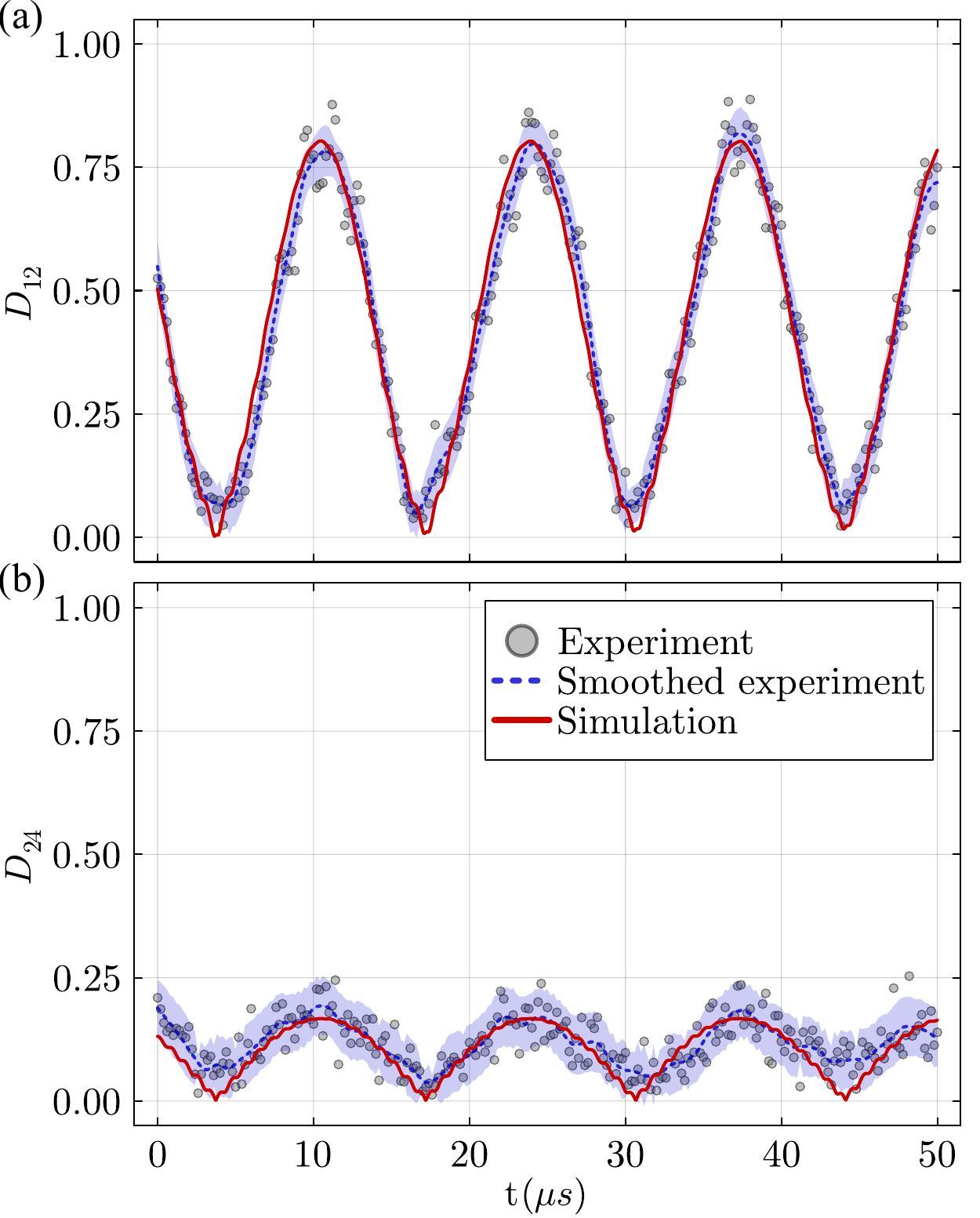}
    \caption{The influence of smoothing on trace distances between trajectories (1, 2) (a) and (2, 4) (b) of the trapped-ion qubit. Figure (b) shows the trace distance with the smallest BLP measure per oscillation. Grey dots represent experimental data, the dotted blue line shows experimental data smoothed with a degree of smoothing s = 0.08, while solid red line represents the results of a numerical simulation with ideal parameters from Table \ref{tab:IonStatesParameters}. Error region for smoothed trace distances shows one standard deviation of 500 shots calculated from experimental data.}
        \label{fig:exampleTD}
\end{figure}

Figure \ref{fig:exampleTD} shows the influence of smoothing with degree $s = 0.08$ on trace distances for the trapped-ion qubit. The dashed blue lines show the smoothed trace distances, while the solid red lines show the results of the numerical simulation with ideal parameters from Table \ref{tab:IonStatesParameters}. The gray dots represent experimental data.

\begin{table}
    \centering
    \begin{tabular}{|c|c|c|c|c|c|c|c|}
    \hline 
        &  1 & 2  & 3 & 4   \\ \hline
$\tau,\, \mu s$    & $3.71$     & $3.71$   & $3.71$  & $4.0$  \\
$\phi$ & $\pi/2$     & $-0.168\pi$   & $3\pi/2$ & $-0.168\pi$   \\
 
\hline
                        
\end{tabular}
    \caption{Parameters of state preparation pulse for trapped-ion experiments. The Rabi frequency of the pulse was set to $\Omega_{\textrm{Rabi}} = 2\pi\times187.5$ kHz.}
    \label{tab:IonStatesParameters}
\end{table}

\subsection{Calculation of statistical errors}

In the measurements of trajectories, each point in time $m$ corresponds to $N$ samples from a Bernoulli distribution with the estimated success probability $p(m)$, with the consequent error:
\beq \label{eq:ProbError}
    \Delta p(m) = \sqrt{\dfrac{p(m)(1-p(m))}{N}}.
\eeq
The corresponding errors for the estimated components of the Bloch vector $v_k(m) = 2p_k(m)-1$, $k \in \{x, y, z \}$ are expressed as:
\beq \label{eq:SigmaError}
    \Delta v_k(m) = \sqrt{\dfrac{1- v_k(m)^2}{N}}.
\eeq

Thus, the error in the trace distance estimation \eqref{eq:TDDecomp} can be calculated as (the dependence on time $m$ is omitted):
\beq \label{eq:TDError}
    \Delta D_{ij} = \frac{1}{2}\sqrt{\sum\limits_{k}\left( \dfrac{\partial D}{\partial v_k}\right)^2(v_{k,i}^2 + v_{k,j}^2)},
\eeq
where
\beq \label{eq:TDDer}
    \left(\dfrac{\partial D}{\partial v_k}\right)^2 = \left(\dfrac{\partial D}{\partial v_{k,i/j}}\right)^2 = \dfrac{(v_{k,i} - v_{k,j})^2}{D_{ij}^2}.
\eeq

\section{Description of the trapped-ion experimental setup}\label{App:Ion}

\begin{figure}
\centering
\includegraphics[width=1.0\textwidth]{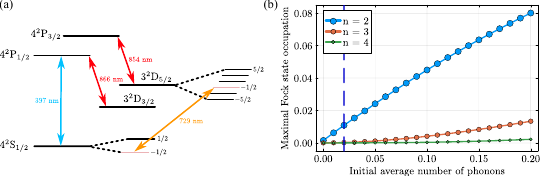}
\caption{(a) Schematic picture of energy levels of $^{40}$Ca$^+$ ion (not to scale). The $\ket{4S_{1/2}, \, m = -1/2}$ and $\ket{3D_{5/2}, \, m = -1/2 }$ Zeeman sublevels are used as the qubit levels $\ket{0}$ and $\ket{1}$ respectively. (b) Maximal occupation number of Fock states with $n = 2, 3, 4$ of a trapped-ion qubit depending on initial average number of phonons. The points are connected by lines for clarity. Dashed vertical line shows the initial average number of phonons $\bar{n} = 0.02$ that was achieved during the measurements in the trapped-ion experiments. We expect the occupation of $n = 2$ and higher levels to be well below $2$ \%, which is below the accuracy of our measurements and can be neglected.}
\label{fig:CaLvls}
\end{figure}

We use a single trapped $^{40}$Ca$^+$ ion in a linear Paul trap. The overview of a trapped-ion qubit can be found in Refs. [39-41]. The qubit degrees of freedom of the $^{40}$Ca$^+$ ion are represented by Zeeman energy sublevels $\ket{4S_{1/2}, \, m = -1/2}$ and $\ket{3D_{5/2}, \, m = -1/2 }$ at the magnetic field of approximately 4 G (see Fig. \ref{fig:CaLvls}(a)). In our setup single ion is located at the center of a linear Paul trap with secular frequencies \{$\omega_x, \omega_y, \omega_z\} = \{1.6, 1.9, 0.591\}$ MHz. We use the combination of Doppler and sideband cooling techniques to reach the number of phonons of $\bar{n}=0.02$. Doppler cooling is achieved with the red-detuned laser at 397 nm that drives the S$_{1/2}$ $\leftrightarrow$ P$_{1/2}$ transition while the blue-detuned repump laser at 866 nm prevents the state leakage to the metastable D$_{3/2}$ state. Manipulations with the quadrupole $\ket{4S_{1/2}, \, m_i}\leftrightarrow\ket{3D_{5/2}, \, m_j}$ transitions are realized with the ULE-locked Ti:Sa laser at 729 nm referred to as the driving laser. For sideband cooling, we drive the red sideband of $\ket{4S_{1/2}, \, m = -1/2}\leftrightarrow\ket{3D_{5/2}, \, m = -5/2}$ transition. The preparation of the $\ket{4S_{1/2}, \, m = -1/2}$ state is carried out by addressing the $\ket{4S_{1/2}, \, m = 1/2}\leftrightarrow\ket{3D_{5/2}, \, m = -3/2}$ transition. In order to shorten the lifetime of metastable D$_{5/2}$ state during both sideband cooling and state preparation protocols, the repump lasers at 854 nm and 866 nm are applied.

In our setup the driving laser beam is directed along the trap axis. It coincides with the trap axial direction; thus, below we consider the oscillation only along the $z$ axis with the frequency $\omega$, which represents the motional degree of freedom. Quantization of the ion motion along the $z$ axis is described as: 
\beq \label{eq:NMode}
    \hat{z} = \sqrt{\dfrac{\hbar}{2m\omega}} \left (a^\dagger + a \right ),
\eeq
where $\hat{z}$ is the quantized axial coordinate of the ion, $\hbar$ is Planck's constant, $m$ is the mass of the ion, $a^\dagger/a$ are creation/annihilation operators of the motional excitation (called phonons).

The driving laser is considered as a classical monochromatic plane wave acting at resonance with the qubit transition. Hamiltonian of a single ion interacting with the laser in a frame rotating with laser frequency $\omega_{\text{laser}}$ reads (here we use the rotating wave approximation):
\beq \label{eq:IonHamFull}
    H =  \hbar \omega a^\dagger a + \hbar \frac{\Omega}{2} \left [e^{i\eta(a^\dagger+a )-i\varphi} \sigma^++h.c. \right ],
\eeq
where $\Omega, \phi$ are the Rabi frequency and the phase of the qubit laser interacting with the qubit transition, respectively, $\eta = \dfrac{\omega_{laser}}{c}\sqrt{\dfrac{\hbar}{2m\omega}}$ is the Lamb-Dicke parameter with $c$ the speed of light. Figure \ref{fig:CaLvls}(b) shows that for the low initial number of phonons $\bar{n}=0.02$ achieved in the experiments, the maximal occupation of the Fock levels with $n\geq2$ does not exceed $2\,\%$ during the system time evolution, which is below the accuracy of our experiments. The motional degree of freedom in that case can be limited to only two low-lying levels with high accuracy. This validates the qubit approximation for the ion motion used in this work. Hence, in the Hamiltonian \eqref{eq:IonHamFull} the quantum harmonic oscillator creation/annihilation operators are replaced by qubit $\sigma_\pm^R$ operators. After expansion to the first order in the Lamb-Dicke parameter $\eta$ it becomes:
\beq\label{eq:IonHam}
    H = \hbar \frac{\Omega}{2} \sigma_x + \hbar \dfrac{\omega}{2} \sigma_z^R  + \hbar \frac{\Omega \, \eta}{2} \, \sigma_y\sigma_x^R.
\eeq
Moving to a new basis using the following unitary transformation:
\beq\label{eq:IonUnitary}   \sigma_x \rightarrow \sigma_z,  \quad
 \sigma_y \rightarrow \sigma_x,  \quad  \sigma_z \rightarrow \sigma_y,  \eeq
and using the rotating wave approximation, one obtains Eq. (3) of the main manuscript with the system-environment coupling constant $g = \Omega \,\eta $.

\section{Numerical verification of regularization procedure}

\begin{figure}
     \centering
     \includegraphics[width=0.5\linewidth]{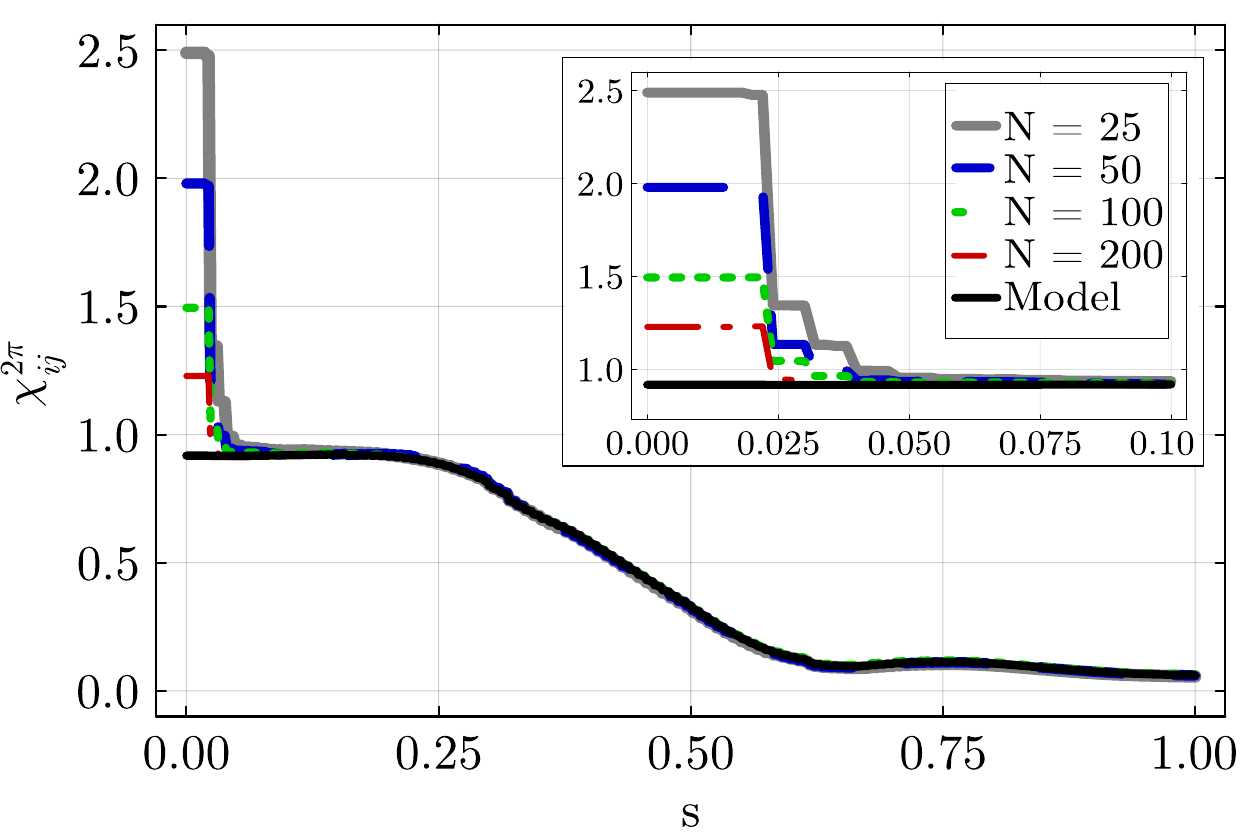}
     \caption{Regularization of the state-pair-dependent non-Markovianity BLP measure per oscillation for numerically simulated data. The solid black line is the simulation without errors. Other colored lines with different styles represent the measure calculated for data with different simulated error rates. The error at each point in time of the simulated trace distance is distributed according to Binomial distribution with number of trials $N$ that is shown on legend. The larger the $N$, the smaller the error rate and the lower the initial noise peak of the measure at $s = 0$.}
     \label{fig:simulatedIonChiOnS}
\end{figure}

To verify the regularization procedure, we numerically simulated the regularization of the BLP measure for the modeled noise. Figure \ref{fig:simulatedIonChiOnS} shows the results for binomially distributed errors with the number of trials $N$. At each point of time $m$ the value of trace distance $D_{noisy}(m)$ is sampled from the binomial distribution with the number of trials $N$ and the success probability given by the numerically simulated value $D(m)$ for the pair of trapped ion initial states (1, 3). The noisy trace distances are then smoothed with variable degrees of smoothing for different numbers of trials $N$. The solid black line in Fig. \ref{fig:simulatedIonChiOnS} shows the results of ideally simulated data without errors. Each curve with errors has a starting region with values exceeding the theoretical value of the BLP measure per oscillation and reaches a plateau which coincides with the plateau for the curve without errors. This behavior is the same as for the experimental results of the trapped-ion qubit. On the plateau, the statistical errors are removed, while physical oscillations are not affected. The greater the $N$, the smaller the statistical error and the smaller the starting noise peak of the non-Markovianity measure for zero degree of smoothing.